# Neural Correlates of Face Familiarity Perception

*Evan Ehrenberg[1], Kleovoulos Leo Tsourides[2], Hossein Nejati[3],*
*Ngai-Man Cheung[3], Pawan Sinha[1*]*

[1]Department of Brain and Cognitive Sciences
Massachusetts Institute of Technology
Cambridge, MA 02139

[2]Department of Computation and Neural Systems
California Institute of Technology
Pasadena, CA 91125

[3]Singapore University of Technology and Design
Singapore

*Corresponding author

**In the domain of face recognition, there exists a puzzling timing discrepancy between results from macaque neurophysiology on the one hand and human electrophysiology on the other. Single unit recordings in macaques have demonstrated face identity specific responses in extra-striate visual cortex within 100 milliseconds of stimulus onset. In EEG and MEG experiments with humans, however, a consistent distinction between neural activity corresponding to unfamiliar and familiar faces has been reported to emerge around 250 ms. This points to the possibility that there may be a hitherto undiscovered early correlate of face familiarity perception in human electrophysiological traces. We report here a successful search for such a correlate in dense MEG recordings using pattern classification techniques. Our analyses reveal markers of face familiarity as early as 85 ms after stimulus onset. Low-level attributes of the images, such as luminance and color distributions, are unable to account for this early emerging response difference. These results help reconcile human and macaque data, and provide clues regarding neural mechanisms underlying familiar face perception.**

The ability to quickly categorize faces as familiar or unfamiliar is a fundamental building block of social cognition in primates. Underscoring the ecological significance of face processing, past research has found evidence for cortical regions in primate brains that are sensitive to face stimuli (Bentin et al., 1996; Bötzel & Grüsser, 1989; Desimone, 1991; Kanwisher et al., 1997; Pinsk et al., 2005; Sams et al., 1997; Tsao et al., 2003; Tsao et al., 2008). Much of this work has demonstrated differentiation in neural activity when viewing faces versus non-face objects. Significant progress has recently been made in identifying neural correlates of the familiar versus unfamiliar face distinction (Castello et al., 2017; Ramon and Gobbini, 2017). However, results from human electrophysiology studies (Caharel et al., 2002; Liu et al., 2002; Tanaka et al., 2006; Ambrus et al., 2019) and non-human primate neurophysiology (Oram & Perrett, 1992; Perrett et al., 1987) reveal an inconsistency related to timing of responses.

Neural activity relating to face identity, assessed via single-unit recordings in macaque superior temporal sulcus (STS), has been observed as early as 70-90 ms post-stimulus onset (Perrett et al., 1987). Even accounting for the known temporal differences in monkey and human brains, potentially due to different anatomical sizes (Fried et al., 2014), this result suggests that face-familiarity driven distinctions should be evident in neural activity of the human brain by 110 ms (Proctor & Brosnan, 2013). Thorpe et al.'s findings showing ultra-fast image classification with eye saccades as a 'read-out' also suggest that neural activity relating to facial familiarity should be evident by this time (Crouzet et al., 2010; Kirchner & Thorpe, 2006; Thorpe et al., 1996). The human neural electrophysiology data thus far, however, belie this expectation.

MEG and EEG components corresponding to the basic face/non-face distinction in human studies have been reported at latencies of approximately 170 ms, after visual stimulus onset (Bentin et al., 1996; Bötzel & Grüsser, 1989; Liu et al., 2002; Sams et al., 1997). Components related to perceived familiarity are observed significantly later, at 250, 400, and 600 ms. These are referred to as the N250, N400 and P600 markers for familiarity (Bentin & Deouell, 2000; Eimer, 2000; Schweinberger et al., 2002; Tanaka et al., 2006). Recent work on rapid saccades to familiar faces (Castello and Gobbini, 2015), also suggests that there must be neural responses well-before the reported familiarity related ERP components. Figure 1 summarizes the latency results by placing several key findings on a common temporal axis. The source of the discrepancy between latency data of familiarity responses arising from macaque studies on the one hand and human studies on the other is unclear.

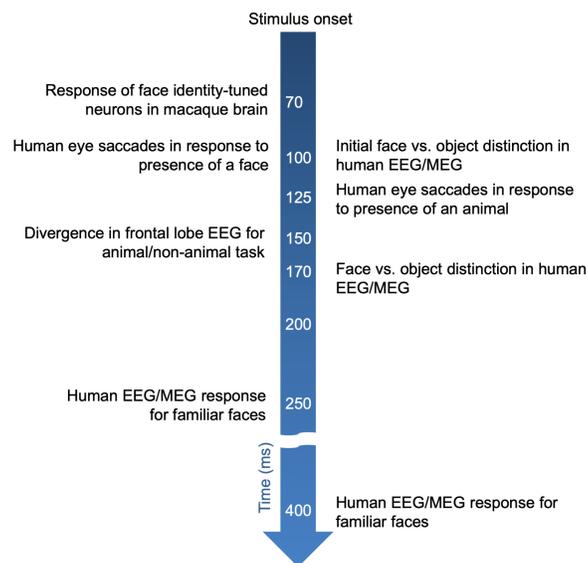

*Figure 1.* A summary of results on the timing of object and face specific responses in human and non-human primate subjects. Of interest to us here is the discrepancy between the early face-identity specific responses observed in macaques (around 70 ms post-stimulus onset) and the relatively late EEG/MEG correlates of face familiarity in humans (250 ms and beyond). Note also the early emergence of behavioral saccades (100-125 ms) in humans for categorizing images.

Given the well-established macaque and human brain homologies and, specifically, conserved structure of the visual pathways, the proposition of there being fundamentally different mechanisms for face processing in the two species, with markedly different latencies, seems unlikely. Let us consider three possibilities that do not critically rely upon positing major species-specific differences. The first of these is that notwithstanding the rapid response at the level of individual neurons, large-scale neural activity needed for a perceptual judgment, and observable via EEG/MEG, may take much longer to emerge and become manifest (Barragan-Jason et al., 2013). Another possibility, however, is that neuronal correlates of face familiarity may indeed arise early, consistent with the results from macaque as well as ultra-rapid saccade discrimination tasks in humans, but have so far not been discovered in the complex electrophysiological data recorded in human EEG/MEG experiments. The second possibility is the more fundamental of the two; if it holds, it obviates need for the first. The third possibility is that the familiar face information starts early and builds up until a person can actually to use the signal to perceive facial familiarity at a later time.

The analytical techniques used in conventional EEG/MEG studies are limited in their ability to mine complex electrophysiological signals. Foremost among such techniques is evoked response analysis (Galambos & Sheatz, 1962; Srinivasan, 2007). This involves averaging time-locked signals over many trials and multiple sensors to discern shared components such as the N170, P600 and others. This pooling of data across trials and space smears much of the fine temporal resolution of raw electrophysiological recordings, and compromises the already modest spatial resolution of EEG/MEG. Furthermore, evoked response analysis is not well suited to discerning spatio-temporally distributed neural patterns that may potentially be the correlates of a specific perceptual judgment.

In order to detect neural signatures of familiarity, which may be too subtle for evoked response analyses, the work we report here uses pattern classification techniques. These techniques are not only more powerful for high-dimensional data mining (Caruana et al., 2008) since they are sensitive to distributed processing, rather than being reliant on large spikes in clustered sensors, they also reduce biases in the analyses by being agnostic about where in the spatial array of sensors, and when in time, relevant information might reside. Combining these analytical techniques with an experimental front-end that uses a stimulus set controlled for various low-level factors, gives us an opportunity to determine whether there may be early neural correlates of the perceptual distinction between familiar and unfamiliar faces.

Recent investigations that have used such techniques have reported encouraging results. Using medium density EEG recordings (32 sensors), Barragan-Jason et al. (2015) have reported finding familiarity-related modulations of the N170 ERP component, with pattern analysis yielding similar results. Using a similar setup, but with a more constrained stimulus set (four identities, two of each gender), Ambrus et al. (2019) have found that identity across different genders begins to be discriminable around 140 ms after stimulus onset, but more robust discrimination becomes evident around 400 ms. In the work we report here, we build upon this work in a few significant ways. First, we use high-density MEG recordings to enhance our ability to detect familiarity related information in the neural signals. The higher sensor density, coupled with the improvement in SNR that MEG provides over EEG for focal activations (Goldenholz et al., 2009), may improve detection of familiarity related signals. Second, we included stimuli besides familiar and unfamiliar faces to be able to determine whether any discrimination observed are driven by face-familiarity or a domain-general familiarity assessment. We also incorporated a condition with person names to test whether the observed distinction might be driven by high-level person familiarity rather than the visual appearance of faces. Additionally, we compare discrimination accuracy and time-course of the face versus non-face distinction relative to the familiar face versus unfamiliar face distinction to be able to directly assess whether these two processes are contemporaneous or temporally staged. The results, as we describe below, reveal early occurring correlates of familiarity judgments specific to faces and help align human electrophysiology data with findings from macaque neurophysiology.

Sixteen adult subjects participated in our study. While in an MEG scanner, they were shown 520 images in random order. Our stimulus set, synopsized in figure 2, included familiar and unfamiliar faces matched in terms of their low-level attributes such as brightness and size, and also with respect to dimensions such as gender, age and attractiveness. We additionally included familiar and unfamiliar objects in our stimulus set in order to test whether the observed response was specific to faces or a generalized familiarity response. If we were to find the same pattern of activity for both familiar faces and familiar objects, it would weaken the case for its consideration as a correlate of facial familiarity. Finally, we included familiar and unfamiliar names on a skin-tone ellipse (matching average size, luminance and color characteristics of the face stimuli used) in order to determine whether the response was specific to the visual percept of faces, rather than to a higher-level cognitive concept of the recognized individual. If we found activity that responded to both familiar faces and familiar names (such as the activity displayed by the so-called 'concept' cells (Quiroga et al., 2005)) this would indicate a generalized response to a person's identity rather than a response driven by visual face recognition.

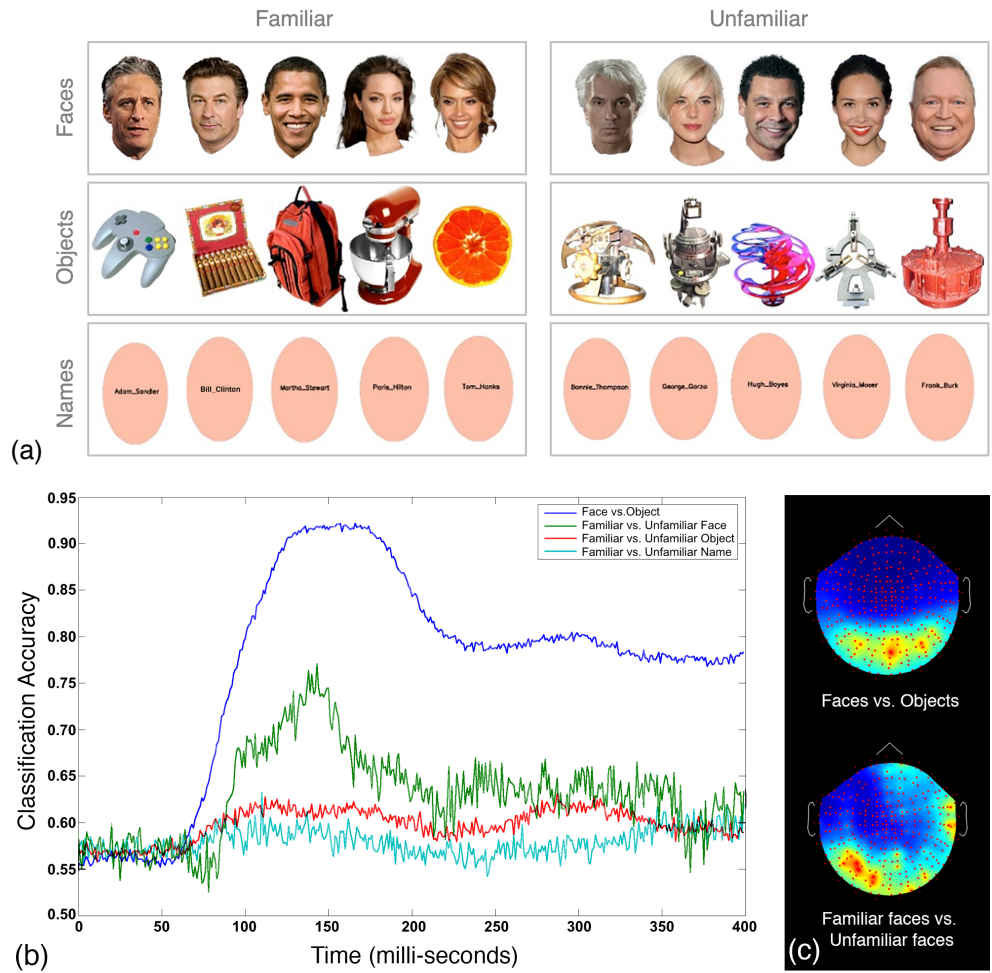

*Figure 2.* (a) Examples of stimuli used in this study. There were six categories: Familiar and unfamiliar faces, familiar and unfamiliar objects, names of familiar and unfamiliar individuals on skin-toned ovals. (b) Classification performance for differentiating between Faces vs. Objects (blue), Familiar vs. Unfamiliar faces (green), Familiar vs. Unfamiliar Objects (red), and Familiar vs. Unfamiliar Names (cyan). Performance trained on a randomly labeled data set represents chance performance. (c) Distribution of sensor weights from the classifier showing the spatial distribution of information contributing to the face/object distinction (upper panel) and familiar/unfamiliar face distinction (lower panel) at 145 ms post stimulus-onset.

The stimulus images were displayed passively with no task other than a requirement to maintain fixation at the center of the screen. Images were each displayed once for 300 ms with random inter-stimulus delay intervals ranging from 1000 ms to 1500 ms. After the MEG scan session, a questionnaire was administered to subjects to assess their familiarity with each of the stimuli.

For analyzing the recorded brain activity traces, we trained a GLMNET logistic regressor (Friedman et al., 2010; Qian et al., 2013) on our data to search for neural activity patterns relating to familiarity. The classifier learns differences between the classes within the MEG data (for example, familiar faces and unfamiliar faces), and is able to utilize distributed patterns and timing differences in the data. Since anatomical variations in sulci and gyri across different brains cause significant differences in the magnetic fields at the scalp (Ahlfors et al., 2010), averaging across subjects runs the risk of obliterating valuable individual-specific patterns. To address this concern, our technique trains a classifier on individual trials from the experiment, and then tests the model's performance on data from other trials from the same subject. Assessing classifier performance on individual trials is a stringent test of robustness, as MEG data are typically noisy. Indeed, most ERF analyses have to average across hundreds of trials, in order to be able to reduce the effects of noise and distinguish between the signals evoked by two different stimulus types. In contrast to this conventional approach, we train and test on single trials from individual subjects with no averaging across instances.

## Results

Following the pre-processing steps of noise modeling and outlier removal (detailed in the Methods section), we subjected the MEG data to a classification analysis. The classifier was provided signals from all magnetometers within 50 ms wide windows. By sliding this window forward one millisecond at a time, we could accurately pinpoint precisely when a significant difference starts to appear between neural activity evoked by two classes. Figure 2b summarizes the results. The face versus object distinction is the first to emerge. Classification performance rapidly improves to being significantly above chance by 70 ms, reaching a maximum of 90% by 120 ms. This timeline is well in advance of the conventionally accepted one, wherein weak face/non-face discrimination is first reported around 100 ms and stronger distinction is seen at approximately 170 ms (Bentin et al., 1996a; Bötzel & Grüsser, 1989; Liu et al., 2002; Sams et al., 1997).

While this evidence of rapid face/non-face distinction is interesting, of greater significance to us is the familiar versus unfamiliar face discrimination. Our pattern classification analyses reveal a clear and early emerging distinction in the neural activity corresponding to these two classes. As is evident in figure 2b, performance rises to significantly above chance level by 85 ms, and eventually attains a maximum of over 75% by 145 ms. Notably, the remaining two classification tasks of discriminating between familiar and unfamiliar objects, and between familiar and unfamiliar names did not yield significant performance (red and cyan curves respectively in figure 2b). This indicates that the distinction we have found is specific to visual face familiarity; it is not a generalized familiarity response or a correlate of high-level person familiarity.

Beyond establishing the existence of a neural correlate of facial familiarity, we can also gain insight into the spatial locations of the sensors that carry information about this distinction. We do this by inspecting the model created by the classifier, and localizing the sensors weighted most strongly (a measure of their discrimination power) to reach the observed level of performance. This method of investigating sensor activity allows us not only to trace manifest brain activity (the magnetic field changes across the scalp), it also allows us to identify *informative* activity specific to the discrimination task at hand. Figure 2c shows the sensor localization results for the face/object as well as the familiar/unfamiliar face discrimination tasks at 145 ms when classification accuracy is highest. Notice that information for the face/object distinction appears to be available largely in the occipital region, while the familiar/unfamiliar face distinction incorporates information from the right-temporal region.

Figure 3 shows the temporal evolution of informative activity corresponding to familiar/unfamiliar face distinction. It is interesting to note that clusters of informative sensors are evident over the early visual cortex by 85 milliseconds, and this locus then smoothly shifts to the right temporal lobe implicating it in this subtle perceptual judgment.

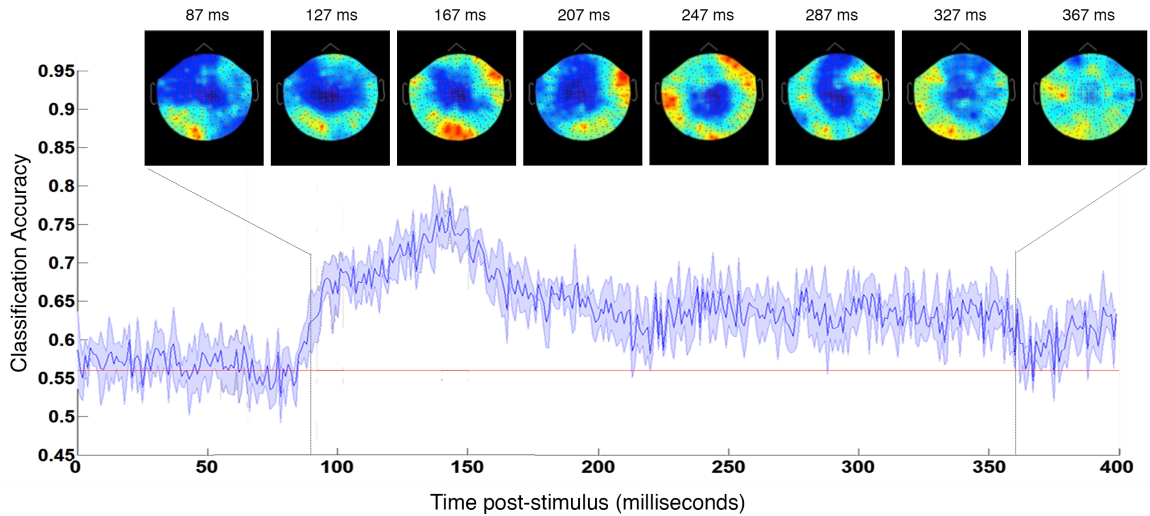

*Figure 3.* *A sequence of heat-maps that display the weights assigned by the familiar vs. unfamiliar face regression to 306 different magnetometer sensors as time progresses after stimulus onset. The snapshot sequence begins at 87 ms, 2 ms after we begin to see significant classification. We can see how the neural activity which is most useful for classifying familiar vs. unfamiliar faces spreads from early occipital cortex pre-100 ms, towards a strong right temporal activation when we near the N170 and N250 time points. This activation appears to then continue to spread towards the frontal lobe before dissipating and becoming less consistently grouped across trials and subjects.*

## Discussion

We applied a two-class logistic regressor to MEG data obtained from subjects viewing images of familiar and unfamiliar faces, objects, and names. We were able to successfully discriminate between brain activity evoked by a face or object with over 90% accuracy. More significantly, our pattern classification analyses revealed systematic differences between activity patterns evoked by familiar and unfamiliar faces; the two patterns could be discriminated robustly with 78% classification accuracy. Importantly, the data revealed evidence of early emergence of these distinctions in the neural traces; significantly above-chance classification accuracy was evident as early as 85 ms for familiar vs. unfamiliar face classification, and 70 ms for face vs. object classification.

These findings of early occurring neural correlates of facial familiarity perception bridge the temporal gap between identity sensitive face responses in monkeys and humans. In doing so, the results not only argue for similarity of face processing neural mechanisms across the two species, but also help account for rapid visual processing tasks such as those performed by participants in studies by Thorpe and others (Crouzet et al., 2010; Kirchner & Thorpe, 2006; Thorpe et al., 1996; VanRullen & Thorpe, 2001). In the light of these early neural markers for face familiarity, we can revisit past EEG and MEG work on face discrimination. The M100 and N250 markers, which have been thought to signify the first neural responses to faces and facial familiarity respectively, are delayed by 30-150 ms beyond the results presented here. The delay leads us to conclude that these markers likely do not represent the earliest feed-forward visually driven responses, but rather larger-scale consolidation of activity perhaps also incorporating feedback activity that may reflect some cognitive influences.

These results provide clues regarding the mechanisms that subserve face-familiarity judgments. First, past work has highlighted the open issue of whether the same mechanisms subserve face/non-face judgments and face individuation (Grill-Spector & Kanwisher, 2005). Given the temporal and spatial distinctions in our data between the two tasks, we infer that the underlying mechanisms are likely not the same. Second, following the logic of Thorpe and colleagues (Thorpe et al., 1996), the early onset of familiarity specific activity evident in our data suggests that judgments of face familiarity are perhaps based largely on feed-forward processing. In order for neural activity to propagate to the visual cortical areas where we find familiarity related information in such a short period of time, and assuming rate coding with neurons that have maximal firing rates of around 100 Hz, only a few neuronal stages can be traversed. This leaves little room for top-down processing and points to the usefulness of a temporally efficient spike coding approach, such as Thorpe's Rank Order Coding algorithm (Thorpe et al., 2001).

Given the likelihood of bottom-up processing underlying our results, we have to ask whether the very rapid distinction we have observed between familiar and unfamiliar faces could be driven by low-level differences in image attributes across the classes (Crouzet & Thorpe, 2011; Thierry et al., 2007). In order to examine this possibility, we sought to classify images in our stimulus set using image descriptors popular in the computational vision community, including SIFT (Lowe, 2004), SURF (Bay et al., 2008), Gabors (Jain et al., 1997), 2D-FFTs (Jain, 1989) and color histograms (Van De Sande et al., 2010). The results are summarized in figure 4. Interestingly, while the face/object distinction can be accomplished via low-level features, the familiar/unfamiliar face discrimination remains at chance for all descriptor types used, suggesting that differences in low-level attributes across the two classes cannot adequately explain the rapidity with which a neural distinction emerges between familiar and unfamiliar faces.

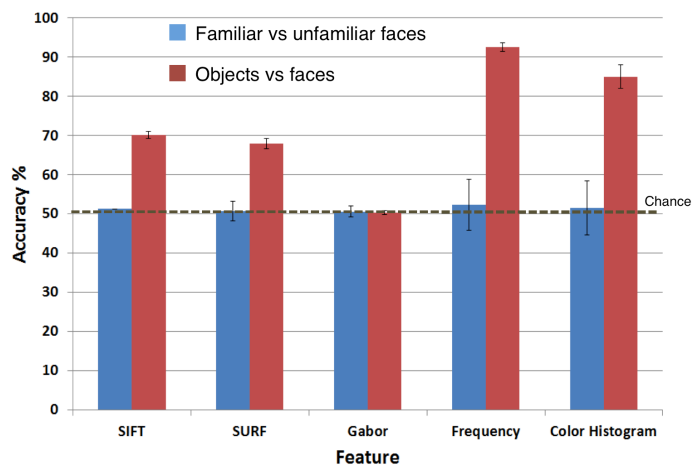

*Figure 4.* Classifier performance while distinguishing image classes using pixel values from the images, rather than evoked brain activity. Comparisons are made across various methods of image pre-processing and feature extraction. While the face/non-face distinction can potentially be explained by low-level attribute differences in the corresponding images (red bars), the familiar/unfamiliar face classes are statistically indistinguishable irrespective of the representational options we examined (blue bars).

Turning to the spatial aspect of our data, we see an overall progression of neural activity from occipital to infero-temporal cortex. This progression is more circumscribed for face detection; activity is seen to stay confined to anterior and middle occipito-temporal cortical regions for this task. However, when comparing this with activation from familiar face processing, we see that activity moves further forward in the right temporal lobe: towards the right superior temporal sulcus in anterior temporal lobe, congruent with recording locations in work showing early onset face identity specific neural responses in macaques (Perrett et al., 1987). That we see activation of both temporal lobes for face detection, but a much stronger activation in the right temporal lobe for face discrimination also agrees with work showing that the left temporal lobe gradually increases in activation with the face-ness of a stimulus, whereas the right temporal lobe operates more categorically, responding strongly when a stimulus is almost definitely a face (Meng et al., 2012), suggesting that the right temporal lobe is involved in more concrete, sub-class classification such as familiarity and identification.

Methodologically, this work adds to the repertoire of analytical tools used in human electrophysiology. The techniques we have used are able to achieve high classification accuracy despite testing on individual trials, obviating the need for signal averaging across multiple trials. This is notable given the susceptibility of the subtle signal differences engendered by facial familiarity to be masked by the noise inherent in scalp recorded electrophysiological signals. Furthermore, the sensor weights generated by the classifier provide a principled way to perform a more impartial ERF analysis, using the sensors deemed important by the model.

The ability to look at a snapshot of neural activity from an individual trial and determine with 90% classification accuracy if the individual was viewing a face or an object, and with 78% accuracy about whether that face was familiar or unfamiliar, both image classes having the same brightness and on-screen size, highlights the usefulness of unbiased pattern classification techniques for analyzing electrophysiological data. This ability to classify with high accuracy on a single trial basis allows for unique applications in scenarios requiring the assessment of a person's familiarity with an individual where continued exposure to an unfamiliar face for ERF averaging may inadvertently familiarize the subject with the target face.

In future studies, it may be interesting to investigate whether the extent of familiarity with an individual (say, personally familiar faces versus celebrity faces) affects the classifier's accuracy, timing of significant discriminatory neural activation, and location of sensors used in the classifier's discrimination. Going beyond familiarity, the methodology used here could also potentially help identify neural correlates of other perceptual judgments such as facial emotions, age or gender.

In summary, the finding of neural markers for face familiarity as early as 85 milliseconds bridges the temporal gap that has persisted in human electrophysiology and macaque single unit recording studies. These markers also hold the promise of helping probe the nature of visual attributes that serve as the drivers of facial familiarity and identity, and associated questions regarding the time-course of familiarization with an initially unfamiliar face.

# Methods

*A. Scanner*

Recordings were made using an Elekta magnetoencephalography scanner. Signals were pre-processed with Brainstorm software and exported to MATLAB for further analysis using our own scripts. Pre-processing steps are described further in section E.

*B. Subjects*

16 adults (6 men and 10 women, age-range: 18-40 years, mean-age: 25 years) participated in our study. All had normal or corrected-to-normal visual acuity and none had a history of neurological or psychiatric disorders. Each subject gave written informed consent according to procedures approved by the MIT IRB.

*C. Experimental Setup*

Subjects passively viewed images of familiar and unfamiliar faces, objects, and names while seated in the MEG scanner. Stimuli were displayed in random order for 300 ms each with a inter-stimulus delay of 1,000-1,500 ms consisting of a central white fixation cross. 40 different familiar celebrity faces were shown, each with 5 different photos, resulting in 200 images of familiar celebrity faces. 40 different unfamiliar foreign celebrity faces were also shown, each with 5 different photos for 200 images of foreign unfamiliar celebrity faces. Subjects underwent a behavioral task after the scan where they labeled the degree of familiarity they had with the various faces (scale 1 to 5, 1 being certain they do not know them and 5 being they can name the celebrity and their occupation). Subjects who failed to provide a familiarity score of 4 or higher to at least 35 of the familiar faces, or who failed to provide a familiarity score of 2 or lower to at least 35 of the unfamiliar faces were removed from further analysis. The familiar and unfamiliar object image classes each had 40 distinct images (one image of each object). Familiar objects were chosen as house-hold or otherwise common objects that subjects would easily recognize, and unfamiliar objects consisted of obscure tools and abstract art; items which are clearly objects and not entirely foreign, and yet are not nameable save by an expert. For names there were 20 familiar American celebrity names taken from 20 of the 40 American celebrity faces used, and 20 pseudo-generic unfamiliar American names were used for the unfamiliar name class.

*D. Image Normalization*

Images were cropped from the background, and normalized to have the same average luminance across image classes. Face images were rotated so that the eyes were aligned horizontally and interpupillary distance was 92-98 pixels. Object images were resized to be of comparable size to the faces on an image-by-image basis. Names were displayed on a skin tone ellipse similar to the face images in an attempt to maintain similar low-level visual features across classes. Face and object images were tested using the GLMNET classifier with various pre-processing and filters, as well as advanced feature extraction using SIFT and SURF image descriptors.

*E. Classifier-based Analysis*

We use Stanford University GLMNET implementation of two-class logistic regressor (Qian et al., 2013) as the core of our pattern classification techniques.

The specific data pre-processing steps, in their order of application, are as follows:

1. Subtract the mean value of the baseline.

2. Filter the signal using a low pass 50 Hz equi-ripple linear 51 taps long filter (technically a band pass .5 Hz -50 Hz).

3. Normalize the data by first computing the standard deviation of the 100ms baseline signal (prior to stimulus onset) and using the result to compute the z-score for the rest of the MEG signal.

4. Create a sliding window of 50ms and vectorize the data.

5. Model the noise using the resting state (i.e. 100ms before the stimulus presentation).

6. Remove outlier trials

In the first step of our preprocessing, we first subtract the mean value of the baseline. This traditional procedure in MEG/EEG analysis serves as one of the basic noise reduction techniques. To further reduce the noise and also remove the signal drift, we apply 0.5-50Hz band-pass filter. Next we normalize the MEG data by calculating the Z-Scores of the entire matrix of 306 MEG sensors $\times$ 1100 ms, separately for each trial from each subject. Calculating the Z-score for the entire signal, while separating trials, provides an unbiased normalization, without loss of data signatures in single trials. After these initial pre-processing steps, we need to slice the data into temporal windows and then format each window into a vector for classification. Based on our empirical results, we selected 50ms as the window width to balance between classifier performance and the algorithm run time. Thus, at this step, the input to the classifier is a 15,300 (50ms time-points$\times$306 sensors) long vector per trial. We refer to each of these vectors as a sample. Each sample is then labeled as one of two classes according to its respective stimuli (familiar faces, unfamiliar faces, objects, names, etc. depending on the classification task). Finally, we divide the entire sample set into two non-overlapping training and testing sets, using 4-fold cross validation. We train the GLMNET classifier on the training set, and then test its performance on the testing set. We apply this method for all four classification tasks, namely, Face vs. Objects, Familiar vs. Unfamiliar Faces, Familiar vs. Unfamiliar Objects, and Familiar vs. Unfamiliar Names.

In order to more accurately model the signal noise, we use the baseline signal as the representation of the noise in our system. The baseline contains no data related to the stimuli, as it represents neural activity generated prior to stimulus presentation. However, the baseline signal is not just a random signal; it represents resting-state activity in the brain. We input the first 50 milliseconds of the baseline in the same format as the other samples to the classifier, to train the classifier on the noise. Any sample from the non-baseline data, that is classified as baseline would be removed. After removal of these samples, we perform the same train-test scenario on the remaining data.

The final step of our pre-processing is to remove small numbers of outliers among the samples. Reviewing the supervised clustering of the data, we discovered a few samples in some of the subjects that were not clustering with the rest of the trials in the same category. Therefore, we allowed up to 10% outlier removal in our classification.

After signal pre-processing, we trained and tested the linear regression with the normalized data. We do not use a validation set; instead, we choose the best model according to the performance (best AUC) on the test set. This allows us to acquire more descriptive models. Of course, using a validation set will cause a slight degradation in the reported performance and is more indicative of the goodness of generalization of the model. We investigate each subject separately such that a 4-fold cross-validation is performed on all trials recorded from the corresponding subject. We then choose the best model according to the performance (best AUC) on the training set, and we apply this model to classify the testing set. We report the average of classification performance and use the best model for sensor localization. We retrieved the classifier weights of each MEG sensor from the trained classification model, and, based on these weights, localized the sensor bearing the highest level of information about the classification task (e.g. face familiarity).